\documentclass[12pt]{article}
\usepackage{epsfig}
\usepackage{amsmath, multirow, amssymb, mathrsfs, wasysym, gensymb}
\usepackage{hyperref}
\newcommand{\JFKslashed}{/ \hspace{-0.2cm}}

\def\beq{\begin{equation}}
\def\eeq#1{\label{#1}\end{equation}}
\def\eeqn{\end{equation}}

\def\beqa{\begin{eqnarray}}
\def\eeqa#1{\label{#1}\end{eqnarray}}
\def\eeqan{\end{eqnarray}}

\let\bar=\overbar

\def\Dslash{\not{\hbox{\kern-4pt $D$}}}
\def\dslash{\not{\hbox{\kern-2pt $\del$}}}

\def\msb{{\bar{\ssstyle M \kern -1pt S}}}

\def\Title#1{\begin{center} {\Large {\bf #1} } \end{center}}

\begin{document}

\Title{Direct CP violation in non-leptonic charm decays: Implications for new physics}

\bigskip\bigskip


\begin{raggedright}  

{\it Jernej F. Kamenik\index{Kamenik, J.F.}\\
Institut Jo\v zef Stefan\\
Jamova 39, P. O. Box 3000\\
SI-1001 Ljubljana, SLOVENIA}
\bigskip\bigskip

{\it Proceedings of CKM 2012, the 7th International Workshop on the CKM Unitarity Triangle, University of Cincinnati, USA, 28 September - 2 October 2012}

\end{raggedright}

\section{Introduction}

CP violation in charm provides a unique probe of New Physics (NP). Not only is it sensitive to NP in the up-quark sector, in the Standard Model (SM) charm processes are dominated by two generation physics with no hard GIM breaking, and thus CP conserving to first approximation. 


CP violation in neutral $D$ meson decays to CP eigenstates $f$ is probed with  time-integrated CP asymmetries $(a_f)$. These can arise from interferences between decay amplitudes with non-zero CP odd ($\phi_f$) and even ($\delta_f$) phase differences  
$a_f^{\rm dir} \simeq  -{2 {r_f} \sin \delta_f \sin \phi_f}$\,,
where $r_f\ll 1$ is the ratio of magnitudes of the interfering amplitudes.  
Recently both the LHCb~\cite{JFKlhcb} and CDF~\cite{JFKCDF10784} collaborations reported evidence for a non-zero
value of the difference $\Delta a_{CP} \equiv a_{K^+ K^-} - a_{\pi^+ \pi^-}$. Combined with other measurements of these CP asymmetries~\cite{JFKHFAG}, the present world average is
\begin{equation}
\Delta a_{CP} = -(0.68\pm 0.15)\%\,.
\label{JFK_eq:acpExp}
\end{equation}

This observation calls for a reexamination of theoretical expectations within the SM. 
Using CKM unitarity ($\sum_{q=d,s,b}\lambda_q = 0$, where $\lambda_q \equiv V_{cq}V^*_{uq}$), the relevant $D^0 \to K^+ K^-, \pi^+ \pi^-$ decay amplitudes ($A_{K,\pi}$) can be written compactly as
$A_{K,\pi} = \lambda_{s,d}(A_{K,\pi}^{s,d} - A_{K,\pi}^{d,s}) - \lambda_b  A_{K,\pi}^{d,s}$, where $A^q_{K,\pi}$ denote the amplitudes carrying $\lambda_q$ CKM prefactors, and we have neglected the small $A^b_{K,\pi}$ contributions. In the isospin limit the two different isospin amplitudes in the first term provide the necessary condition for non-zero $\delta_{K,\pi}$, while $\phi^{\rm SM}_{K,\pi} = {\rm Arg}(\lambda_b/\lambda_{s,d}) \approx \pm 70^\circ$. On the other hand $r_{K,\pi}$ are controlled by  the CKM ratio $\xi= |\lambda_b/\lambda_s| \simeq |\lambda_b/\lambda_d| \approx 0.0007$. Parametrizing the remaining unknown hadronic amplitude ratios as $R_{K,\pi}^{\rm SM} \equiv -A_{K,\pi}^{d,s} / (A_{K,\pi}^{s,d} - A_{K,\pi}^{d,s})$, the SM contribution to $\Delta a_{CP}$ can be written as
\begin{equation}
\Delta a_{CP} \approx (0.13\%) {\rm Im} (\Delta R^{\rm SM})\,,
\end{equation}
where $\Delta R^{\rm SM} = R_K^{\rm SM} + R_\pi^{\rm SM}$\,. 
One observes that $|{\rm Im}(\Delta R^{\rm SM})| = \mathcal O (2-5) $ is needed to reproduce the experimental results in Eq.~\eqref{JFK_eq:acpExp}\,, in contrast to perturbative estimates in the heavy charm quark limit ($|R_{K,\pi}|\sim \alpha_s(m_c)/\pi \sim 0.1$) (see~\cite{JFKGrossman:2006jg,JFKotherSM1}). However, $\xi$ suppressed amplitudes in the numerator of $R_i$ cannot be constrained by rate measurements alone, and it has been pointed out a long time ago that ``$\Delta I=1/2$ rule" type enhancements are possible~\cite{JFKGolden:1989qx} (see also~\cite{JFKotherSM}). Recently~\cite{JFKBrod:2011re}, an explicit model estimate of potentially large $1/m_c$ suppressed contributions has been performed, yielding $\Delta a_{CP}^{\rm SM}$ $\lesssim 0.4~\%$\,. Although this is an order of magnitude above na\"ive expectations, the experimental value in Eq.~\eqref{JFK_eq:acpExp} cannot be reached.

\section{Implications of $\Delta a_{CP}$ for Physics Beyond SM}

In the following we will therefore assume the SM does not saturate the experimental value, leaving room for potential NP contributions. These can again be parametrized in terms of an effective Hamiltonian valid below the $W$ and top mass scales
\begin{equation}\label{JFK_eq:HNP}
\mathcal H^{\rm eff-NP}_{|\Delta c|=1} = \frac{G_F}{\sqrt 2} \sum_i C_i^{\rm NP(\prime)} \mathcal Q_i^{(\prime)}\,,
\end{equation}
where the relevant operators $\mathcal Q_i^{(\prime)}$ have been defined in~\cite{JFKDCPVD}.
Introducing also the NP hadronic amplitude ratios as $R^{{\rm NP},i}_{K,\pi} \equiv G_F \langle{K^+ K^-, \pi^+ \pi^-}|\mathcal Q^{(\prime)}_i|{D^0}\rangle / \sqrt 2 (A_{K,\pi}^{s,d} - A_{K,\pi}^{d,s})$ and writing $C_i^{\rm NP} = v_{\rm EW}^2 / \Lambda^2$, the relevant NP scale $\Lambda$ is given by
\begin{equation}
\frac{(10~{\rm TeV})^2}{\Lambda^2} = \frac{(0.63\pm 0.14)-0.12 {\rm Im}(\Delta R^{\rm SM})}{{\rm Im}(\Delta R^{{\rm NP},i})}\,.
\end{equation}

Comparing this estimate to the much higher effective scales probed by CP violating observables in $D$ mixing and also in the kaon sector, one first needs to verify, if such large contributions can still be allowed by other flavor constraints. Within the effective theory approach, this can be estimated via so-called ``weak mixing" of the effective operators.
In particular, time-ordered correlators of $\mathcal H^{\rm eff-NP}_{|\Delta c|=1}$ with the SM effective weak Hamiltonian can, at the one weak loop order, induce important contributions to CP violation in both $D$ meson mixing and kaon decays ($\epsilon'/\epsilon$). On the other hand, analogue correlators, quadratic in $\mathcal H^{\rm eff-NP}_{|\Delta c|=1}$ turn out to be either chirally suppressed and thus negligible, or yield quadratically divergent contributions, which are thus highly sensitive to particular UV completions of the effective theory~\cite{JFKDCPVD}. 

\subsection{Universality of CP Violation in $\Delta F=1$ processes}

The strongest bounds can be derived for a particular class of operators, which transform non-trivially only under the $SU(3)_Q$ subgroup of the global SM quark flavor symmetry $\mathcal G_F = SU(3)_Q \times SU(3)_U \times SU(3)_D$, respected by the SM gauge interactions. In particular one can prove that their CP violating contributions to $\Delta F=1$ processes have to be approximately universal between the up and down sectors~\cite{JFKUCPV}. Within the SM one can identify two unique sources of $SU(3)_Q$ breaking given by $\mathcal{A}_u \equiv (Y_u Y_u^\dagger)_{\JFKslashed{\mathrm{tr}}}$ and $\mathcal{A}_d \equiv (Y_d Y_d^\dagger)_{\JFKslashed{\mathrm{tr}}}$, where
$\JFKslashed{\mathrm{tr}}$ denotes the traceless part. Then in the two generation limit,  one can construct a single source of CP violation, given by $J\equiv i [\mathcal{A}_u,\mathcal{A}_d]$~\cite{JFKarXiv:1002.0778}. The crucial observation is that $J$ is invariant under $SO(2)$ rotations between the $\mathcal{A}_u$ and $\mathcal{A}_d$ eigenbases. Introducing now $SU(2)_Q$ breaking NP effective operator contributions  of the form $\mathcal  Q_L = \Big[(X_L)^{ij}\, \overline Q_i \gamma^\mu Q_j \Big] L_\mu$, where $L_\mu$ denotes a flavor singlet current, it follows that their CP violating contributions have to be proportional to $J$ and thus invariant under flavor rotations. The universality of CP violation induced by $\mathcal Q_L$ can be expressed explicitly as~\cite{JFKUCPV}
\begin{equation}\label{Uni2gen}
{\rm Im}(X^u_L)_{12} = {\rm Im}(X^d_L)_{12}\propto {\rm Tr} \left( X_L \cdot
J\right)\, .
\end{equation}

The above identity holds to a very good approximation even in the three-generation framework. In the SM, large values of $Y_{b,t}$ induce a $SU(3)/SU(2)$ flavor symmetry breaking pattern~\cite{JFKKagan:2009bn} which allows to decompose $X_L$ under the residual $SU(2)$ in a well defined way.  Finally, residual SM $SU(2)_Q$ breaking is necessarily suppressed by small mass ratios $m_{c,s}/m_{t,b}$, and small CKM mixing angles $\theta_{13}$ and $\theta_{23}$. 

The most relevant implication of Eq.~\eqref{Uni2gen} is that it predicts a direct correspondence between $SU(3)_Q$ breaking NP contributions to $\Delta a_{CP}$  and $\epsilon'/\epsilon$~\cite{JFKUCPV}. It follows immediately that stringent limits on possible NP contributions to the later, require $SU(3)_Q$ breaking contributions to the former to be below the per mille level (for $\Delta R^{{\rm NP},i}=\mathcal O (1)$).

~

The viability of the remaining 4-quark operators in $\mathcal H^{\rm eff-NP}_{|\Delta c|=1}$ as explanations of the $\Delta a_{CP}$ value in Eq.~\eqref{JFK_eq:acpExp},  depends crucially on their flavor and chiral structure. In particular, operators involving purely right-handed quarks are unconstrained in the effective theory analysis but may be subject to severe constraints from their UV sensitive contributions to $D$ mixing observables.  On the other hand, QED and QCD dipole operators are at present only weakly constrained by nuclear EDMs and thus present the best candidates to address the $\Delta a_{CP}$ puzzle~\cite{JFKDCPVD}.

\section{Explanations of $\Delta a_{CP}$ within NP Models}

Since the announcement of the LHCb result,  several prospective explanations of  $\Delta a_{CP}$ within various NP frameworks have appeared in the literature. In the following we briefly discuss $\Delta a_{CP}$ within some of the well-motivated beyond SM contexts. 

In the Minimal Supersymmetric SM (MSSM), the right size of the QCD dipole operator contributions  can be generated with non-zero left-right up-type squark mixing contributions $(\delta^u_{12})_{LR}$~\cite{JFKGrossman:2006jg,JFKGiudice:2012qq,JFKHiller:2012wf}.
 Parametrically, such effects in $\Delta a_{CP}$ can be written as~\cite{JFKGiudice:2012qq}
\begin{equation}
|\Delta a_{CP}^{\rm SUSY}| \approx 0.6\% \times \left( \frac{|{\rm Im}(\delta_{12}^u)_{LR}|}{10^{-3}} \right) \left( \frac{\rm TeV}{\tilde m} \right)\,,
\end{equation}
where $\tilde m$ denotes a common squark and gluino mass scale. At the same time dangerous contributions to $D$ mixing observables are chirally suppressed.  It turns out however that even the apparently small $(\delta^u_{12})_{LR}$ value required implies a highly nontrivial flavor structure of the UV theory. In particular, large trilinear ($A$) terms and sizable mixing among the first two generation squarks ($\theta_{12}$) are required~\cite{JFKGiudice:2012qq} 
\begin{eqnarray}
{\rm Im}(\delta^u_{12})_{LR} &\approx& \frac{{\rm Im}(A) \theta_{12} m_c}{\tilde m} \approx \left(\frac{{\rm Im}(A)}{3}\right) \left(\frac{\theta_{12}}{0.3}\right) \left(\frac{TeV}{\tilde m}\right) 0.5 \times 10^{-3}\,.\nonumber
\end{eqnarray}

Similarly, warped extra dimensional models~\cite{JFKRS1} that explain the quark spectrum through flavor anarchy~\cite{JFKRS1,JFKRS2} can naturally give rise to QCD dipole contributions 
affecting $\Delta a_{CP}$~\cite{JFKRS}. The dominant one-loop Higgs exchange contribution yields
\begin{equation}
|\Delta a_{CP}^{\rm RS}| \approx 0.6\% \times \left( \frac{\mathcal O_\beta}{0.1} \right) \left( \frac{Y_5}{4} \right)^2 \left( \frac{3 \rm TeV}{m_{KK}} \right)^2\,,
\end{equation}
where $m_{KK}$ is the KK scale, $Y_5$ is the 5D Yukawa coupling in appropriate units of the AdS curvature and the function $\mathcal O_\beta$ parameterizes the Higgs profile overlap with
the fermion KK state wavefunctions.
Reproducing the experimental value of $\Delta a_{CP}$ requires near-maximal 5D Yukawa coupling, 
close to its perturbative bound~\cite{JFKAgashe:2008uz} of $4\pi/\sqrt{N_{KK}} \simeq 7$ for $N_{KK} = 3$ perturbative KK states. In turn, this helps to suppress dangerous tree-level contributions to CP violation in $D-\bar D$ mixing~\cite{JFKDDRS}. This scenario can also be interpreted within the framework of partial compositeness in four dimensions, but generic composite models typically require smaller Yukawas to explain $\Delta a_{C P}$ and consequently predict sizable contributions to CP violation in $\Delta F = 2$ processes~\cite{JFKcomp}.  


\section{Prospects}

Continuous progress in lattice QCD methods ({\it c.f.}~\cite{JFKHansen:2012tf}) gives hope that ultimately the role of SM long distance dynamics in $\Delta a_{CP}$ could be studied from first principles. In the meantime it is important to identify possible experimental tests able to distinguish standard vs. non-standard explanations of the observed value.

Explanations of $\Delta a_{CP}$ via NP contributions to the QCD dipole operators generically predict sizable effects in radiative charm decays~\cite{JFKradiative}. First, in most explicit NP models the short-distance contributions to QCD and EM dipoles are expected to be similar. Moreover, even assuming that only a non-vanishing QCD dipole  is generated at some high scale, the mixing of the two operators under the QCD renormalization group implies comparable size of the two contributions at the charm scale. Unfortunately, the resulting effects in the rates of radiative $D\to X \gamma$ decays are typically more than two orders of magnitude below the long-distance dominated SM effects~\cite{JFKRS}. This suppression can be partly lifted when considering CP asymmetries in exclusive $D^0\to P^+P^-\gamma$ transitions, where $M_{PP} = \sqrt{(p_{P^+}+p_{P^-})^2}$ is close to the $\rho,\omega,\phi$ masses~\cite{JFKradiative} (see also~\cite{Lyon:2012fk}). Related observables in rare semileptonic $D$ decays have also been studied recently~\cite{JFKFajfer:2012nr}.

An alternative strategy makes use of (sum rules of) CP asymmetries in various hadronic D decays (necessarily including neutral mesons). It is effective in isolating possible non-standard contributions to $\Delta a_{CP}$ if they are generated by effective operators with a $\Delta I = 3/2$ isospin structure~\cite{JFKGrossman:2012eb} (which unfortunately does not include the QCD dipoles).


Finally, correlations of non-standard contributions to $\Delta a_{CP}$ with other CP violating observables like electric dipole moments, rare top decays or down-quark phenomenology are potentially quite constraining but very NP model-dependent~\cite{JFKGiudice:2012qq,JFKAltmannshofer:2012ur}.

\bigskip
The author would like to thank the organizers of {\it CKM 2012} for the invitation to this fruitful workshop. This work is supported in part by the Slovenian Research Agency.

\end{document}